\documentclass[prd,preprint,nofootinbib,showpacs,preprintnumbers,amsmath,amssymb]{revtex4}

\usepackage{hyperref}
\hypersetup{
    colorlinks,%
    citecolor=blue,%
    filecolor=blue,%
    linkcolor=blue,%
    urlcolor=blue
}
\usepackage{bm}

\newcommand{\be}{\begin{equation}}
\newcommand{\ee}{\end{equation}}
\newcommand{\bea}{\begin{eqnarray}}
\newcommand{\eea}{\end{eqnarray}}
\newcommand{\p}{\partial}

\begin{document}

\title{Anomalies and Hawking fluxes from the black holes of topologically massive gravity}
\author{Achilleas P. Porfyriadis}
\email{apporfyr@mit.edu}
\affiliation{Department of Physics\\ Massachusetts Institute of Technology\\ Cambridge, MA 02139, USA}
\date{\today}

\begin{abstract}
The anomaly cancelation method proposed by Wilczek et al. is applied to the black holes of topologically massive gravity (TMG) and topologically massive gravito-electrodynamics (TMGE). Thus the Hawking temperature and fluxes of the ACL and ACGL black holes are found. The Hawking temperatures obtained agree with the surface gravity formula. Both black holes are rotating and this gives rise to appropriate terms in the effective U(1) gauge field of the reduced (1+1)-dimensional theory. It is found that the terms in this U(1) gauge field correspond exactly to the correct angular velocities on the horizon of both black holes as well as the correct electrostatic potential of the ACGL black hole. So the results for the Hawking fluxes derived here from the anomaly cancelation method, are in complete agreement with the ones obtained from integrating the Planck distribution.
\end{abstract}


\maketitle

\section{Introduction}

Hawking radiation is perhaps the most prominent effect of quantum gravity. There are various derivations, including Hawking's original one \cite{Hawking:1974sw} calculating Bogoliubov coefficients, which have shown that the radiation is thermal with a blackbody spectrum at a temperature given by the surface gravity, $T_H=\kappa/{2\pi}$. Derivations have been found also using Euclidean quantum gravity \cite{Gibbons:1976ue} and in string theory \cite{Strominger:1996sh}. A derivation using semiclassical methods (WKB) in a tunneling picture has been proposed by Parikh and Wilczek in \cite{Parikh:1999mf} (see also \cite{Berezin:1999nn}).

Recently, a new method for calculating Hawking radiation from the horizon of black holes has been developed by Robinson and Wilczek \cite{Robinson:2005pd} and by Iso, Umetsu, and Wilczek \cite{Iso:2006wa,Iso:2006ut}. The new method ties the existence of Hawking radiation to quantum anomalies at the horizon of black holes. More precisely, it is argued that gauge and gravitational anomalies at the horizon can account for the Hawking fluxes of the black hole. The key step, the one that needs to be carried out separately for each black hole case, is the dimensional reduction of the scalar field action in the black hole background considered in the method. That is, one considers a complex scalar field in the black hole background and reduces its action near the horizon to a collection of $1+1$ dimensional fields. Then using the ideas and calculations in \cite{Robinson:2005pd} one obtains the Hawking temperature $T_H$ from the anomalies cancelation viewpoint. Furthermore, using the analysis carried out in \cite{Iso:2006wa,Iso:2006ut} one may also find the specific Hawking fluxes of the energy-momentum, angular momentum (if the black hole is rotating) and electric charge (if the black hole is charged).

During recent intense work the method proposed by Wilczek {\it et al.} \cite{Robinson:2005pd,Iso:2006wa,Iso:2006ut} has been successfully applied to various black objects \cite{various black holes}, including the newly obtained black holes of five-dimensional minimal gauged supergravity \cite{Porfyriadis:2008yp}. In the direction of further extending the method, higher spin currents were studied under conformal transformations to obtain the full thermal spectrum of Hawking radiation \cite{higher spin}. An interesting point raised in \cite{Iso:2006wa} was that while it is the consistent form of the anomalies that is canceled, the boundary condition employed at the horizon is the vanishing of the covariant form of the current. Following up, in \cite{covariant anomaly "method"} a variant of the original method was proposed in which only covariant forms of the anomalies are used and all its applications \cite{covariant anomaly "apps"} have confirmed the results obtained in the corresponding papers in \cite{various black holes}. For a discussion of the relative merits of these various approaches see \cite{Umetsu:2008cm}. In this Letter however, as we are primarily concerned on applying the method to new black holes, we will follow the original approach of \cite{Robinson:2005pd,Iso:2006wa,Iso:2006ut} only. For other related work on the subject see also \cite{various}.

In this Letter we apply the anomalies cancelation method to obtain the Hawking temperature and fluxes of black holes of topologically massive gravity \cite{TMG}. Three-dimensional topologically massive gravity (TMG) with a negative cosmological constant, $\Lambda=-1/\ell^2$, is known to admit two types of black hole solutions: the Einstein BTZ black hole \cite{Banados:1992wn} which solves trivially the TMG equations of motion and the recently obtained non-Einstein black hole given in \cite{Bouchareb:2007yx}. For the case of the BTZ black hole accounts on applying the anomalies cancelation method may be found in \cite{RW for BTZ}. We will thus concentrate only on the non-Einstein black holes of TMG. The first solution with a vanishing cosmological constant was recently found in \cite{Moussa:2003fc} and the ACL black hole with negative cosmological constant which we are going to consider in this Letter is given in \cite{Bouchareb:2007yx}. We will also analyze the case of the ACGL black hole \cite{Moussa:2008sj} which is the corresponding newly obtained solution in topologically massive gravito-electrodynamics (TMGE).

The rest of the Letter is as follows. In Section \ref{Section: Black body radiation} we derive the fluxes of Hawking radiation from general charged rotating black holes by integrating the Planck distribution. This will serve as a reference for comparing with our result obtained later using the anomalies cancelation method. In Section \ref{Section: ACL black hole} we apply the anomalies cancelation method, following \cite{Robinson:2005pd,Iso:2006wa,Iso:2006ut}, to the ACL black hole and derive its Hawking temperature and fluxes of energy-momentum tensor and angular momentum. We find agreement of the results with the surface gravity formula for the temperature and the results of section \ref{Section: Black body radiation} for the fluxes. In Section \ref{Section: ACGL black hole} we do the same for the ACGL black hole, also finding its charge flux. The results are again consistent with our expectations. Section \ref{Section: Conslusion} is devoted to conclusions and discussions.

\section{Blackbody Radiation}\label{Section: Black body radiation}

In order to compare with our results from anomalies cancelation later, let us here calculate the Hawking fluxes one obtains from integrating the Planck distribution for a general rotating black hole with a nontrivial electromagnetic background gauge field $A$. The appropriate chemical potentials for fields radiated with an azimuthal angular momentum $m$ and an electric charge $e$ are the horizon angular velocity $\Omega_H$ and the corotating electrostatic potential $\Phi$ respectively. For fermions, the Planck distribution for blackbody radiation moving in the positive $r$ direction at a temperature $T_H$ is given by,
\be
N_{e, m}(\omega)= \frac{1}{e^{(\omega - e\Phi - m\Omega_H)/T_H}+1}.
\ee
We consider only fermions in order to avoid superradiance \cite{Iso:2006ut}. From the above distribution and including the contribution from the antiparticles, we find the following Hawking fluxes of electric charge, angular momentum, and energy-momentum tensor respectively:
\bea
F_Q &=& e \int_0^\infty \frac{d\omega}{2\pi} \left(N_{e,m}(\omega) - N_{-e, -m}(\omega)\right)=\frac{e}{2\pi}\left(e\Phi + m\Omega_H\right), \label{flux el charge} \\
F_a &=& m \int_0^\infty \frac{d\omega}{2\pi} \left(N_{e,m}(\omega) - N_{-e, -m}(\omega)\right)=\frac{m}{2\pi}\left(e\Phi + m\Omega_H\right), \label{flux ang momentum} \\
F_M &=& \int_0^\infty \frac{d\omega}{2\pi}~\omega \left(N_{e,m}(\omega) + N_{-e, -m}(\omega)\right)=\frac{1}{4\pi}\left(e\Phi + m\Omega_H\right)^2+ \frac{\pi}{12}T_H^2. \label{flux e-m tensor}
\eea
It should be emphasized that the above fluxes are so observed by a static observer at fixed radius r close to the horizon: the radiation is purely thermal only \emph{at} the horizon. This emission when propagated to infinity experiences the effective potential due to spatial curvature outside the horizon and thus the spectrum of the radiation observed at infinity is modified to that of a three-dimensional grey body at the Hawking temperature \cite{Robinson:2005pd,Iso:2006wa,Iso:2006ut}. For asymptotically flat space-times it is often said loosely that the Hawking fluxes observed at infinity are thermal \emph{if} one ignores the grey body factor. The black holes we are going to consider in this Letter are not asymptotically flat and hence due to spatial curvature at infinity, the fluxes above do not coincide with the ones observed at infinity \emph{even if} one ignores the grey body factor.\footnote{I am grateful to the anonymous referee for raising the point and to Frank Wilczek and Sean Robinson for helping clarify my understanding on this.}

In the following sections we shall derive the above fluxes at the horizon of ACL and ACGL black holes using the gravitational anomalies cancelation method. We will follow \cite{Robinson:2005pd,Iso:2006wa,Iso:2006ut} which derived the fluxes in the Unruh vacuum and only for an observer \emph{at} the horizon (i.e. we neglect the grey body factor at infinity).

\section{ACL black hole} \label{Section: ACL black hole}

ACL black hole is a rotating black hole solution to three-dimensional topologically massive gravity (TMG) described by the action
\be
I_{TMG}=I_E+I_{CS}
\ee
where $I_E$ is the Einstein action,
\be
I_E=\frac{1}{16\pi G} \int d^3x\sqrt{-g}(R-2\Lambda),
\ee
and $I_{CS}$ is the gravitational Chern-Simons action,
\be
I_{CS}=\frac{1}{32\pi G \mu}\int d^3x\sqrt{-g}\,\varepsilon^{\lambda\mu\nu}\Gamma^{\rho}_{\lambda \sigma}\left(\partial_{\mu}\Gamma^
\sigma_{\rho \nu}+\frac{2}{3}\Gamma^\sigma_{\mu\tau}\Gamma^\tau_{\nu \rho}\right).
\ee
Here $1/\mu$ is the Chern-Simons coupling constant, $\varepsilon^{\mu\nu\rho}$ is the Levi-Civita tensor (tensor density given by $\epsilon/\sqrt{-g}$ with $\epsilon^{012}=+1$), and $G$ is the gravitational constant which may be taken both positive or negative.
The metric of the ACL black hole, in ADM form, is given by \cite{Bouchareb:2007yx}
\be\label{metric ADM form}
ds^2=-N^2 dt^2+r^2(d\varphi+N^\varphi dt)^2+\frac{d\rho^2}{\zeta^2 r^2 N^2},
\ee
where
\bea
r^2&=&\rho^2 +2\omega\rho + \omega^2(1-\beta^2)+\frac{\beta^2\rho_0^2}{1-\beta^2}\label{r},\\
N^2&=&\beta^2\frac{\rho^2-\rho_0^2}{r^2}\label{N},\\
N^\varphi&=&-\frac{\rho+(1-\beta^2)\omega}{r^2}, \label{Nphi}
\eea
and
\be
\beta^{2}=\frac{1}{4}\left( 1-\frac{27\Lambda}{\mu^{2}}\right), \qquad \zeta =\frac{2}{3}\mu.
\ee
Regular black holes with finite temperature are obtained in the range $0< \beta^2 <1, \, \rho_0>0$, and $\omega>-\rho_0/\sqrt{1-\beta^2}\, (\omega\neq\rho_0/(1-\beta^2))$. The two horizons are at $\rho=\pm\rho_0$ with
the outer $\rho=\rho_0$ being the event horizon (henceforth called the horizon). The two parameters $\omega$ and $\rho_0$ are related to the black hole mass and angular momentum.

ACL black holes are so far the only known non-Einstein regular black hole solutions of TMG. They were also obtained (in different coordinates) in \cite{Anninos:2008fx} where they are called ``spacelike stretched black holes.'' These black holes are asymptotic to warped AdS$_3$ and it has been shown that they are quotients of warped AdS$_3$ under a discrete subgroup of the isometry group \cite{Anninos:2008fx}, just as BTZ black holes are quotients of AdS$_3$ \cite{Banados:1992gq}.

Consider a scalar complex field in the ACL background. The free part of the action is
\bea
S=-\int d^3x\sqrt{-g}\,g^{\mu\nu}\p_\mu\phi^*\,\p_\nu\phi&=&\frac{1}{\zeta}\int dt\, d\rho\, d\varphi\, \phi^*\left[-\frac{1}{N^2}\p_t^2+\zeta^2\p_{\rho}\,r^2N^2\p_{\rho}\right. \nonumber \\
&&+\left. \left(\frac{1}{r^2}-\frac{{N^\varphi}^2}{N^2}\right)\p_{\varphi}^2+2\frac{N^\varphi}{N^2}\p_t\p_{\varphi} \right]\phi.
\eea
To study the near horizon theory, $\rho\to\rho_0$, it is most convenient to transform first to a ``tortoise'' like coordinate.
In our case if we first transform to the $\rho_*$ coordinate defined by
\be\label{tortoise}
d\rho=\zeta r N^2 d\rho_*,
\ee
and then take the near horizon limit $\rho\to\rho_0$ (\emph{i.e.} $N^2\to 0$) we obtain,
\be
S\approx \int dt\, d\rho_*\, d\varphi\, r\, \phi^* \left[-\p_t^2+\p_{\rho_*}^2-{N^\varphi}^2\p_{\varphi}^2+ 2N^\varphi\p_t\p_{\varphi} \right]\phi.
\ee
Then by expanding the complex field $\phi$ as
\be\label{expansion}
\phi=\sum_m \phi_m(\rho_*,t)\, e^{i m \varphi},
\ee
we may integrate out the $\varphi$ dependence to find that
\bea
S&\approx& \sum_m 2\pi\int dt\, d\rho_*\, r\, \phi_m^* \left[-\p_t^2+\p_{\rho_*}^2+m^2{N^\varphi}^2+2imN^\varphi\p_t
\right]\phi_m \nonumber \\
&=&\sum_m 2\pi\int dt\, d\rho_*\, r\, \phi_m^* \left[-\left(\p_t-i m N^\varphi \right)^2+\p_{\rho_*}^2\right]\phi_m. \label{ACL reduced action}
\eea
The last implies that the near horizon theory has been reduced to an infinite collection of massless $1+1$ dimensional
fields labeled by the quantum number $m$. Indeed, transforming back to the original $(\rho,t)$ coordinates, to each of the fields $\phi_m$ corresponds the action:
\[
S_m=\int dt\, d\rho\, r\, \phi_m^* \left[-\frac{1}{\zeta r N^2}\left(\p_t-i m N^\varphi \right)^2+\p_{\rho}\, \zeta r N^2 \p_{\rho}\right]\phi_m.
\]
That is, each $\phi_m$ can be considered as a $1+1$ complex scalar field in the backgrounds of the dilaton $\Psi=r$, metric
\be\label{ACL reduced metric}
ds^2=-f(\rho) dt^2+\frac{1}{f(\rho)}d\rho^2\, , \qquad \textrm{where} \qquad f(\rho)=\frac{d\rho}{d\rho_*}=\zeta r N^2
\ee
and $U(1)$ gauge potential
\be\label{ACL effective A}
{\cal A}_t=m N^\varphi, \qquad {\cal A}_\rho=0.
\ee
The effective $U(1)$ gauge symmetry here has its roots in the axial isometry of the ACL black hole in the first place. Also note that any possible mass or interaction terms in the full action would have been trivially suppressed in the near horizon limit too, since they would all be multiplied by a factor $f$ in the tortoise coordinates.

At this point we impose the constraint that the classically irrelevant ingoing modes vanish near the horizon and thus the theory becomes chiral. Hence anomalies arise which, as will be shown, may account for the Hawking fluxes of the black hole.

In particular, it was shown in \cite{Robinson:2005pd} that for a metric of the form
(\ref{ACL reduced metric}), the (purely timelike) anomaly for the energy-momentum tensor at the horizon accounts for a flux of massless blackbody radiation at the temperature $T_H=\frac{1}{4\pi}\p_\rho f\vert_{\rho_o}$. Thus it follows that the Hawking temperature of the ACL black hole is given by
\be
T_H=\frac{1}{4\pi}\p_\rho f\vert_{\rho_o}=\frac{1}{4\pi}\zeta r \p_\rho N^2\vert_{\rho_o}=\frac{\mu\beta^2}{3\pi}\frac{\rho_o}{r_h},
\ee
where $r_h=r(\rho_0)=[\rho_0+(1-\beta^2)\omega]/\sqrt{1-\beta^2}$. This result agrees with the one given in \cite{Bouchareb:2007yx} where the Hawking temperature was calculated as the surface gravity at the horizon
$\kappa=n^\rho \p_\rho N\vert_{\rho_0}$ (where $n^\rho=\sqrt{g^{\rho\rho}}=\zeta r N$) divided by $2\pi$.

In \cite{Iso:2006wa,Iso:2006ut} fluxes of  angular momentum, electric charge (if there is one) and energy-momentum tensor were computed from gauge and gravitational anomalies at the horizon, by deriving and solving the Ward identities with appropriate boundary conditions corresponding to the Unruh vacuum. The boundary conditions employed were vanishing of the radial component of the covariant current at the horizon and also (implicitly) vanishing of ingoing current at radial infinity \cite{Iso:2006ut}. The second amounts to setting the integration constant $k^r$, which arises in solving the Ward identity for the contribution of the ingoing modes $K^r$, equal to zero.

Very recently, a modification of the derivation of the fluxes was proposed in \cite{Morita:2009mt} which does not set the integration constant $k^r$ equal to zero. Instead, working with $k^r$ arbitrary, the fluxes are derived using conformal field theory techniques and a different set of boundary conditions which again correspond to the Unruh vacuum. In particular, it is shown in \cite{Morita:2009mt} that in order to derive the fluxes corresponding to $U(1)$ current one has to consider not only the gauge anomaly but also the chiral anomaly and similarly in order to derive the energy-momentum flux one needs to consider the trace anomaly too (in addition to the gravitational anomaly for the energy-momentum tensor). Then, given the appropriate boundary conditions employed in \cite{Morita:2009mt} the same fluxes as in \cite{Iso:2006ut} are derived for the Unruh vacuum.

Since in this Letter we are primarily interested in the application of the method to new black holes we will not review the details of either derivation but rather use the end results for the flux formulas. It is found that for a metric of the form (\ref{ACL reduced metric}) and with an effective $U(1)$ gauge potential as in (\ref{ACL effective A}) the fluxes of angular momentum and energy-momentum are \cite{Iso:2006ut,Morita:2009mt}
\[
-\frac{m}{2\pi}{\cal A}_t(\rho_0) \qquad \textrm{and} \qquad \frac{1}{4\pi}{\cal A}^2_t(\rho_0)+N^r_t(\rho_0)
\]
respectively, with $N^r_t=({f^{'}}^2+f\,f^{''})/192\pi$. We thus find that the flux of angular momentum for the ACL black hole, corresponding to the gauge anomaly, is
\be\label{ACL flux ang momentum}
-\frac{m}{2\pi}{\cal A}_t(\rho_0)=-\frac{m^2}{2\pi}N^\varphi(\rho_0)=\frac{m^2}{2\pi}\Omega_H,
\ee
where $\Omega_H=-N^\varphi(\rho_0)$ is the angular velocity at the horizon, and the flux of energy-momentum corresponding to the gravitational anomaly is
\be\label{ACL flux e-m tensor}
\frac{1}{4\pi}{\cal A}^2_t(\rho_0)+N^r_t(\rho_0)=\frac{m^2}{4\pi}{N^{\varphi}}(\rho_0)^2+\frac{1}{192\pi}(\p_\rho f)^2\vert_{
\rho_0}=\frac{m^2}{4\pi}\Omega_H^2+\frac{\pi}{12}T_H^2.
\ee
We thus find that the Hawking fluxes derived from anomalies cancelation method for the ACL black hole (\ref{ACL flux ang momentum},\ref{ACL flux e-m tensor}) agree with those calculated from integrating the thermal spectrum in (\ref{flux ang momentum},\ref{flux e-m tensor}), as clearly $\Phi=0$ for the ACL black hole.

\section{ACGL black hole} \label{Section: ACGL black hole}

ACGL black hole is a solution to three-dimensional topologically massive gravito-electrodynamics (TMGE), that is, three-dimensional Einstein-Maxwell theory with both gravitational and electromagnetic Chern-Simons terms. It is described by the action
\be
I_{TMGE}=I_E+I_{M}+I_{CSG}+I_{CSE}
\ee
where $I_E$ and $I_M$ are the Einstein and Maxwell actions respectively,
\bea
I_E&=&\frac{1}{16\pi G} \int d^3x\sqrt{-g}(R-2\Lambda),\\
I_M&=&-\frac{1}{4} \int d^3x \sqrt{-g}\, g^{\mu\nu}g^{\rho\sigma}F_{\mu\rho}F_{\nu\sigma},
\eea
and $I_{CSG}$ and $I_{CSE}$ are the gravitational and electromagnetic Chern-Simons actions respectively,
\bea
I_{CSG}&=&\frac{1}{32\pi G \mu_G}\int d^3x\sqrt{-g}\,\varepsilon^{\lambda\mu\nu}\Gamma^{\rho}_{\lambda \sigma}\left(\partial_{\mu}\Gamma^\sigma_{\rho \nu}+\frac{2}{3}\Gamma^\sigma_{\mu\tau}\Gamma^\tau_{\nu \rho}\right),\\
I_{CSE}&=&\frac{\mu_E}{2} \int d^3x \sqrt{-g}\,\varepsilon^{\mu\nu\rho} A_\mu\partial_\nu A_\rho .
\eea
Here the Chern-Simons coupling constants are $1/\mu_G$ and $\mu_E$. The metric of the ACGL black hole, in ADM form, is given by equations (\ref{metric ADM form}$-$\ref{Nphi}) with \cite{Moussa:2008sj}
\be
\beta^{2}=\frac{1-2\lambda-4\Lambda/\mu_E^2}{2(1-\lambda)}, \qquad \zeta =\mu_E, \qquad \textrm{and} \qquad \lambda=\frac{\mu_E}{2\mu_G}.
\ee
Generic regular black holes with finite temperature are obtained for $0<\beta^2<1$ and $\rho_0>0$.\footnote{We have set the time scale parameter in \cite{Moussa:2008sj} to $\sqrt{c}=1$.} The two horizons are at $\rho=\pm\rho_0$ with the outer $\rho=\rho_0$ being the event horizon (henceforth called the horizon) and the two parameters $\omega$ and $\rho_0$ are related to the black hole mass and angular momentum. Present is also a gauge electromagnetic potential given by
\be\label{ACGL A gauge field}
A=A_t dt+A_\varphi d\varphi,
\ee
where
\be
A_t=\pm\sqrt{\frac{3\lambda-1}{8\pi G}}(1-\beta^2) \qquad \textrm{and} \qquad A_\varphi=\mp\sqrt{\frac{3\lambda-1}{8\pi G}}\left(\rho+(1-\beta^2)\omega\right).
\ee
For this electromagnetic potential to be real the Chern-Simons coupling constants should obey appropriate bounds: if the gravitational constant $G$ is taken to be positive then we should have $\mu_E/\mu_G>2/3$, otherwise if $G$ is negative then $\mu_E/\mu_G<2/3$.

Considering a scalar complex field in the ACGL background we have the free part of the action
\bea
S=-\int d^3x\sqrt{-g}\,g^{\mu\nu}(D_\mu\phi)^*\,D_\nu\phi&=&\frac{1}{\zeta}\int dt\, d\rho\, d\varphi\, \phi^*\left[-\frac{1}{N^2}D_t^2+\zeta^2\p_{\rho}\,r^2N^2\p_{\rho}+\right. \nonumber \\
&+&\left.\left(\frac{1}{r^2}-\frac{{N^\varphi}^2}{N^2}\right)D_{\varphi}^2+2\frac{N^\varphi}{N^2}D_t D_{\varphi} \right]\phi,
\eea
where $D_t=\p_t+ieA_t$ and $D_\varphi=\p_\varphi+ieA_\varphi$ and $e$ is the electric charge of $\phi$. Transforming to the tortoise coordinate (\ref{tortoise}) and taking the near horizon limit we find,
\be
S\approx \int dt\, d\rho_*\, d\varphi\, r\, \phi^* \left[-D_t^2+\p_{\rho_*}^2+{N^\varphi}^2D_{\varphi}^2-2N^\varphi D_t D_{\varphi} \right]\phi.
\ee
Then by expanding the field $\phi$ as in (\ref{expansion}) we may integrate out the $\varphi$ dependence to find that
\be\label{ACGL reduced action}
S\approx \sum_m 2\pi\int dt\, d\rho_*\, r\, \phi_m^* \left[-(D_t-ieA_\varphi N^\varphi-imN^\varphi)^2+\p_{\rho_*}^2
\right]\phi_m.
\ee
The last implies that the near horizon theory has been reduced to an infinite collection of massless $1+1$ dimensional
fields labeled by the quantum number $m$. Indeed, transforming back to the original $(\rho,t)$ coordinates, to each of the fields $\phi_m$ corresponds the action:
\[
S_m=\int dt\, d\rho\, r\, \phi_m^* \left[-\frac{1}{\zeta r N^2}\left(\p_t+ie(A_t-N^\varphi A_\varphi)-imN^\varphi \right)^2+\p_{\rho}\, \zeta r N^2 \p_{\rho}\right]\phi_m.
\]
That is, each $\phi_m$ can be considered as a $1+1$ complex scalar field in the backgrounds of the dilaton $\Psi=r$, metric
\be\label{ACGL reduced metric}
ds^2=-f(\rho) dt^2+\frac{1}{f(\rho)}d\rho^2\, , \qquad \textrm{where} \qquad f(\rho)=\frac{d\rho}{d\rho_*}=\zeta r N^2
\ee
and $U(1)$ gauge potential ${\cal A}_\mu$,
\be\label{ACGL effective A}
{\cal A}_t=-e(A_t-N^\varphi A_\varphi)+m N^\varphi, \qquad {\cal A}_\rho=0.
\ee
The effective $U(1)$ gauge symmetry here consists of two pieces: the first one came from the original electromagnetic gauge field of the ACGL black hole (\ref{ACGL A gauge field}) while the second one is again a result of the axial isometry of the ACGL black hole.

Now ignoring the classically irrelevant ingoing modes near the horizon we will find that the anomalies arising in the chiral theory account for the Hawking fluxes of the black hole. First, following \cite{Robinson:2005pd} for the metric
(\ref{ACGL reduced metric}) the anomaly for the energy-momentum tensor at the horizon gives a flux of massless blackbody radiation at the temperature $T_H=\frac{1}{4\pi}\p_\rho f\vert_{\rho_o}$. Thus it follows that the Hawking temperature of the ACGL black hole is given by
\be
T_H=\frac{1}{4\pi}\p_\rho f\vert_{\rho_o}=\frac{1}{4\pi}\zeta r \p_\rho N^2\vert_{\rho_o}=\frac{\mu_E\beta^2}{2\pi}\frac{\rho_o}{r_h},
\ee
where $r_h=r(\rho_0)=[\rho_0+(1-\beta^2)\omega]/\sqrt{1-\beta^2}$. This result agrees with the one given in \cite{Moussa:2008sj} where the Hawking temperature was calculated as the surface gravity at the horizon
$\kappa=n^\rho \p_\rho N\vert_{\rho_0}$ (where $n^\rho=\sqrt{g^{\rho\rho}}=\zeta r N$) divided by $2\pi$.

Next, we proceed to calculate the fluxes of  electric charge, angular momentum, and energy-momentum tensor
from gauge and gravitational anomalies at the horizon. For a metric of the form (\ref{ACGL reduced metric}) and with an effective $U(1)$ gauge potential as in (\ref{ACGL effective A}) the flux of electric charge is
\bea
-\frac{e}{2\pi}{\cal A}_t(\rho_0)&=&-\frac{e}{2\pi}\left[-e(A_t-N^\varphi A_\varphi)+m N^\varphi\right]\vert_{\rho_0} \nonumber \\ &=&\frac{e}{2\pi}\left[e( A_t(\rho_0)+\Omega_H A_\varphi (\rho_0) )+m\Omega_H\right], \label{ACGL flux el charge}
\eea
where $\Omega_H=-N^\varphi(\rho_0)$ is the angular velocity at the horizon. As the Killing vector of the ACGL black hole is $\xi=\p_t+\Omega_H \p_\varphi$, the corotating electrostatic potential is,
\[
\Phi=\xi^\mu A_\mu \vert_{\rho_0}=A_t(\rho_0)+\Omega_H A_\varphi (\rho_0)
\]
which shows that (\ref{ACGL flux el charge}) agrees with the standard result (\ref{flux el charge}). Similarly, we find for the ACGL black hole, that the flux of angular momentum is
\be\label{ACGL flux ang momentum}
-\frac{m}{2\pi}{\cal A}_t(\rho_0)=\frac{m}{2\pi}\left[e\Phi+m\Omega_H\right],
\ee
and the flux of energy-momentum is
\be\label{ACGL flux e-m tensor}
\frac{1}{4\pi}{\cal A}^2_t(\rho_0)+N^r_t(\rho_0)=\frac{1}{4\pi}\left[e\Phi+m\Omega_H\right]^2+\frac{\pi}{12}T_H^2.
\ee
We note that (\ref{ACGL flux ang momentum}) and (\ref{ACGL flux e-m tensor}) are also seen to agree with the standard results (\ref{flux ang momentum}) and (\ref{flux e-m tensor}) respectively.

\section{Conclusion-Discussion}\label{Section: Conslusion}

In this Letter we have applied the method of anomalies cancelation to derive the Hawking temperature and fluxes of the ACL and ACGL black holes of topologically massive gravity and gravito-electrodynamics. It was shown that near the horizon the quantum field behaves as an infinite set of two-dimensional conformal fields labeled by only one quantum number. The effective two-dimensional theory near the horizon is described by charged matter fields in an electric field. In particular, for both black holes, the azimuthal symmetry lead to a $U(1)$ gauge symmetry for each partial field mode and the respective $U(1)$ charge is given by the corresponding azimuthal quantum number. Upon suppressing classically irrelevant ingoing modes, we have calculated the Hawking temperature and fluxes of the black holes from the gauge and gravitational anomalies that arise in the chiral theory. The results found are consistent with the surface gravity formula and the fluxes obtained from integrating the Planck distribution, respectively.

The method of Wilczek \emph{et al.} adopted here uses only quantum anomalies at the horizon and therefore it is quite universal, in the sense that it does not depend on the details of the quantum fields away from the horizon. The success of the method though crucially relies on the dimensional reduction of the d-dimensional action (for the scalar field used to probe a particular d-dimensional black hole background) to the 2-dimensional ones on the $(\rho,t)$ plane. Therefore, the important equations of this Letter are equations (\ref{ACL reduced action}) and (\ref{ACGL reduced action}) which ensure that the rest of the method will yield fluxes in complete agreement with the Planckian ones. The anomaly method has been applied to a large number of black objects to date [8], but the cases in this Letter are new in that it is the first time black holes from topologically massive gravity are considered using this method.

In general, it is not clear \emph{a priori} for any black hole in any number of dimensions and coming from any theory of gravity that the dimensional reduction for the probing scalar field is at all possible, or even if possible, that it will yield such two-dimensional actions (i.e. with such an $f(\rho)$ and such an effective U(1) gauge field ${\cal A}_t$), that the anomaly cancelation method will give correct Hawking fluxes. Though not well understood why, it seems it is always possible to reduce a scalar field action in the background space-time of a black hole to an infinite sum of two-dimensional conformal field actions near the horizon. Therefore, an important development would be establishing the reduction of the action in general, for every black hole, so that a case by case application is no longer necessary. The ADM form of the black holes in this Letter suggests that such a general reduction may actually be possibly performed for a general ADM metric describing a black hole. Making sure that the metric actually describes a black hole is crucial and amounts to putting certain assumptions on the lapse and the shifts along the lines in \cite{Carlip:1998wz}. Work in that direction is in progress.

\acknowledgments
I would like to thank Frank Wilczek and Sean Robinson for enlightening discussions on the subject.

\newpage

\emph{Note added}: After completion of this work the preprint \cite{Nam:2009dd} appeared in the arXiv which has some overlap with Section \ref{Section: ACL black hole}. In \cite{Nam:2009dd} the authors obtain the Hawking fluxes using the covariant methods of \cite{covariant anomaly "method"} and the results agree with the ones obtained here in section \ref{Section: ACL black hole} following the original method of \cite{Robinson:2005pd,Iso:2006wa,Iso:2006ut}.

\bibliography{References}

\end{document}